\documentclass[aps,prc,reprint,amsmath,amssymb,showpacs,superscriptaddress]{revtex4-1}

\usepackage{CJK}
\usepackage{graphicx}% Include figure files
\usepackage{dcolumn}% Align table columns on decimal point
\usepackage{bm}% bold math
\usepackage{epstopdf}

\usepackage{color}
\usepackage{hyperref}% add hypertext capabilities
\usepackage{threeparttable}
\usepackage{booktabs}
\usepackage{multirow}
\usepackage{longtable}
\usepackage{supertabular}

\begin{document}
\begin{CJK*}{UTF8}{} % Use default fonts from CJK (see below)


\title{The role of near neutron drip-line nuclei in the $r$-process}

\author{T. Yu \CJKfamily{gbsn} (余婷)}
\affiliation{Department of Physics, Fuzhou University, Fuzhou 350108, Fujian, China}

\author{Y. Y. Guo \CJKfamily{gbsn} (郭粤颖)}
\affiliation{National Astronomical Observatories, Chinese Academy of Sciences, Beijing 100101, PR China}
\affiliation{School of Astronomy and Space Science, University of Chinese Academy of Sciences, Beijing 100049, PR China}

\author{X. F. Jiang \CJKfamily{gbsn} (姜晓飞)}
\affiliation{State Key Laboratory of Nuclear Physics and Technology, School of Physics, Peking University, Beijing 100871, China}

\author{X. H. Wu \CJKfamily{gbsn} (吴鑫辉)}
\email{wuxinhui@fzu.edu.cn}
\affiliation{Department of Physics, Fuzhou University, Fuzhou 350108, Fujian, China}

%\thanks{}
%\date{\today}

\begin{abstract}
  The role of near neutron-drip-line nuclei in the rapid neutron-capture process ($r$-process) is studied with the classical $r$-process model.
  Simulations under different astrophysical conditions ($T$, $n_n$) show that $r$-process paths approach the neutron-drip line under low-temperature and high-neutron-density conditions.
  A sensitivity study reveals that variations in the nuclear masses of these exotic nuclei lead to obvious abundance variations in the $A=110-125$, $A=175-185$, $A=200-205$, and superheavy regions.
  By contrast, the $r$-process rare-earth peak and the $A=130,195$ peaks remain largely unaffected.
  The nuclei that obviously impact $r$-process abundances are mainly distributed in the region of $25\leq Z\leq 90$ and $50\leq N\leq 180$, with the nuclei around neutron magic numbers found to be particularly important for the $r$-process, even in the near-neutron-drip-line region.
\end{abstract}

\maketitle

\end{CJK*}


%-----------------------------------------------------
\section{Introduction}\label{secIntr}
%-----------------------------------------------------

The production of about half of the heavy elements found in nature is assigned to a specific astrophysical nucleosynthesis process, i.e., the rapid neutron-capture process ($r$-process).
Although this idea was proposed about seven decades ago~\cite{Burbidge1957Rev.Mod.Phys.}, a full understanding of $r$-process remains one of the most challenging topics due to the respective difficulties in both astrophysics and nuclear physics~\cite{Arnould2007Phys.Rep., Thielemann2011Prog.Part.Nucl.Phys., Thielemann2017Annu.Rev.Nucl.Part.Sci., Kajino2019Prog.Part.Nucl.Phys., Cowan2021Rev.Mod.Phys.}.

The astrophysical sites responsible for the $r$-process abundances are still not fully understood, with some proposed scenarios still under debate.
Ejecta from neutron star mergers (NSMs) is supported to be an $r$-process site by the GRB170817 kilonova~\cite{Abbott2017Astrophys.J., Pian2017Nature, Watson2019Nature} associated with gravitational waves from GW170817~\citep{Abbott2017Phys.Rev.Lett.}; however, its role in producing the $r$-process abundances observed in the oldest metal-poor stars remains debated due to the ``time delay".
The neutrino driven wind (NDW)~\cite{Woosley1994Astrophys.J.} and the magneto-hydrodynamic jet (MHDJ)~\cite{Nishimura2015Astrophys.J., Nishimura2012Phys.Rev.C} from core-collapse supernova (CCSN), as well as the outflows from collapsar~\cite{Siegel2019Nature, Nakamura2015Astron.Astrophys., Famiano2020Astrophys.J.} are favored $r$-process candidate sites that could have occurred in the early Universe. 
This makes them suitable for studies of nuclear cosmochronology~\cite{Hill2017Astron.Astrophys., Wu2022Astrophys.J.152, Wu2023Sci.Bull., Huang2025ApJ}.
However, there are debates on whether desired high-entropy conditions to produce actinides can occur in the NDW from CCSN~\cite{Fischer2010Astron.Astrophys.}.
The rapid time scale of the MHDJ from CCSN and the outflows from collapsar would lead to the underproduction of isotopic abundances above and below the main $r$-process peaks~\cite{Kajino2019Prog.Part.Nucl.Phys.}, which are different from the observed $r$-process abundances.

The $r$-process nucleosynthesis path in the nuclear chart runs to the very neutron-rich region, where presently only limited experimental information is available, thus relying heavily on theoretical predictions of nuclear properties.
For extreme $r$-process astrophysical conditions with lower electron fraction $Y_e$ (higher neutron fraction) or lower density, the $r$-process path may even run close to the neutron-drip line, i.e., boundary of bound nuclei.
The study of exotic nuclei near the neutron-drip line is therefore important not only for satisfying humanity's curiosity to understand exotic nuclear structures but also for understanding the origin of elements in nature.
It is thus important to know (1) under what specific astrophysical conditions the nuclear properties of exotic nuclei near the neutron-drip line play an important role in the $r$-process and (2) how the $r$-process abundances are affected by uncertainties in the properties of these exotic nuclei.

In this work, the $r$-process conditions that lead to $r$-process paths close to the neutron-drip line are studied with the classical $r$-process model.
Under these typical conditions, the impacts of uncertainties in the nuclear masses of these exotic nuclei near the neutron-drip line on the $r$-process abundances are analyzed.

\section{Classical $R$-process Model}

Based on the situation that astrophysical sites responsible for the $r$-process abundances are still not fully understood, we perform $r$-process simulations based on a site-independent $r$-process model, the so-called classical $r$-process model~\cite{Kratz1993Astrophys.J., Kratz2007ApJ, Sun2008Phys.Rev.C, Xu2013Phys.Rev.C, Zhao2019Astrophys.J., Jiang2021Astrophys.J.}, which can be regarded as a simplification of the dynamical $r$-process model, and has been successfully employed in describing $r$-process patterns of both the solar system and metal-poor stars.

In the classical $r$-process model, iron group seed nuclei are irradiated by high-density neutron sources with a high temperature $T \gtrsim 1.5~\mathrm{GK}$.
The $r$-process abundances are obtained by the superposition of abundances from the simulations in several neutron flows with different neutron densities, temperatures, and irradiation times.
The weights can be determined by fitting to the solar $r$-process abundances.

In the astrophysical environments with high-temperature $T \gtrsim 1.5~\mathrm{GK}$ and high neutron density $n_n \gtrsim 10^{20} \mathrm{~cm}^{-3}$, the equilibrium between neutron capture and photodisintegration reactions can be achieved, and the abundance ratios of neighboring isotopes on an isotopic chain can be obtained by the Saha equation~\cite{Cowan1991Phys.Rep., Qian2003PPNP, Arnould2007Phys.Rep.}
\begin{equation}\label{Eq:saha}
\begin{aligned}
  \frac{Y(Z, A+1)}{Y(Z, A)}= &n_n\left(\frac{2 \pi \hbar^2}{m_\mu k T}\right)^{3 / 2} \frac{G(Z, A+1)}{2 G(Z, A)}\left(\frac{A+1}{A}\right)^{3/2} \\
  &\times \exp \left[\frac{S_n(Z, A+1)}{k T}\right],
\end{aligned}
\end{equation}
where $Y(Z, A), S_n(Z, A)$, and $G(Z, A)$ are, respectively, the abundance, one-neutron separation energy, and partition function of nuclide $(Z, A)$, and $\hbar$, $k$, and $m_\mu$ are the Planck constant, Boltzmann constant, and atomic mass unit, respectively. 
Note that the neutron separation energy $S_n$ deduced from nuclear masses appears in the exponential, suggesting the importance of nuclear masses in the equilibrium.
Note that here in the classical $r$-process model, the $r$-process path is defined as the path that goes through the most abundantly distributed nucleus in each isotopic chain, as determined by Eq.~\eqref{Eq:saha}.

The abundance flow from one isotopic chain to the next is governed by $\beta$ decays and can be expressed by a set of differential equations
\begin{equation}\label{Eq:beta}
\begin{aligned}
  \frac{d Y(Z)}{d t}= & Y(Z-1) \sum_A P(Z-1, A) \lambda_\beta^{Z-1, A} \\
  &-Y(Z) \sum_A P(Z, A) \lambda_\beta^{Z, A},
\end{aligned}
\end{equation}
where $\lambda_\beta^{Z, A}$ is the $\beta$-decay rate of the nucleus $(Z,A)$, $Y(Z)=\sum_A Y(Z$, $A)=\sum_A P(Z, A) Y(Z)$ is the total abundance of each isotopic chain, $ P(Z, A)$ is the individual population coefficients. 
By using Eqs.~\eqref{Eq:saha} and \eqref{Eq:beta}, the abundance of each isotope can be determined. 
After the neutrons freeze-out, the unstable isotopes on the neutron-rich side will proceed to the stable isotopes mainly via $\beta$ decays, and the final abundances are obtained.

The nuclear physics inputs for the classical $r$-process model involve nuclear masses and $\beta$-decay half lives.
The nuclear masses are taken from one of the WS4 nuclear mass model~\cite{Wang2014Phys.Lett.B}, if the
experimental data~\cite{Wang2021Chin.Phys.C} are not available.
For the $\beta$-decay rates, the empirical formula~\cite{Zhou2017Sci.ChinaPhys.Mech.Astron.} is adopted with decay energies from the adopted nuclear masses if the experimental data~\cite{NNDC} are not available.

\section{$R$-process conditions for yielding neutron-drip-line $R$-process paths}

\begin{figure*}[!htbp]
  \centering
  \includegraphics[width=0.7\textwidth]{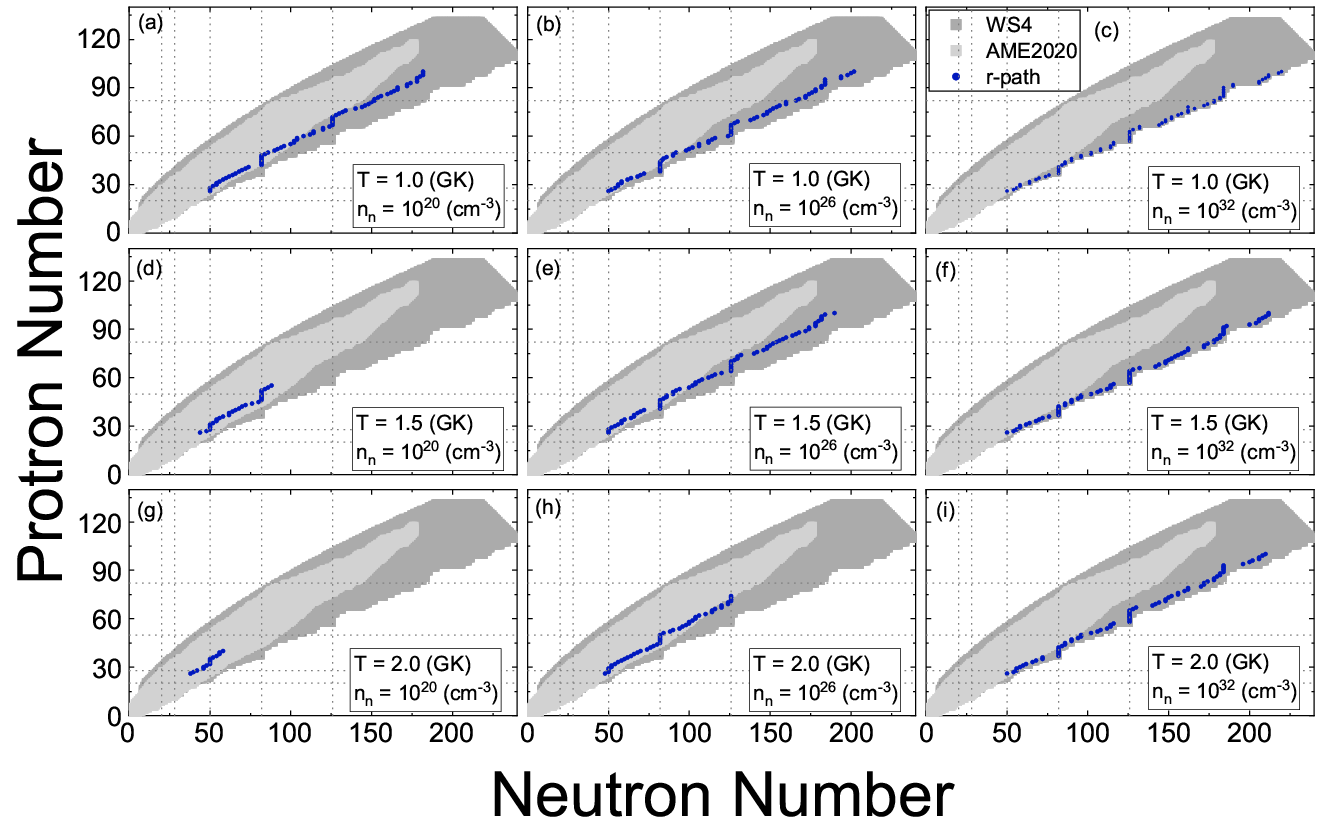}
  \caption{$r$-process nucleosynthesis paths under different astrophysical conditions, characterized by temperature ($T$) and neutron number density  ($n_n$).}
\label{fig1}
\end{figure*}

The first step is to identify the astrophysical conditions that cause the $r$-process nucleosynthesis paths to go close to the neutron-drip line.
The $r$-process paths under different astrophysical conditions $(T,n_n)$ are presented in Fig.~\ref{fig1}.
It can be seen that the $r$-process paths are closer to neutron-drip line for both higher neutron density $n_n$ and lower temperature $T$.
This can be easily understood with Eq.~\eqref{Eq:saha}, as both higher neutron density $n_n$ and lower temperature $T$ would lead to a larger ratio of $Y(Z,A+1)/Y(Z,A)$.
This thus leads to a more extensive abundance distribution on the neutron-rich side, i.e., $r$-process paths shift toward the neutron-rich region.

%----------------------------------------------------------------------------------------
%----------------------------------------------------------------------------------------

\begin{figure*}[!htbp]
  \centering
  \includegraphics[width=0.7\textwidth]{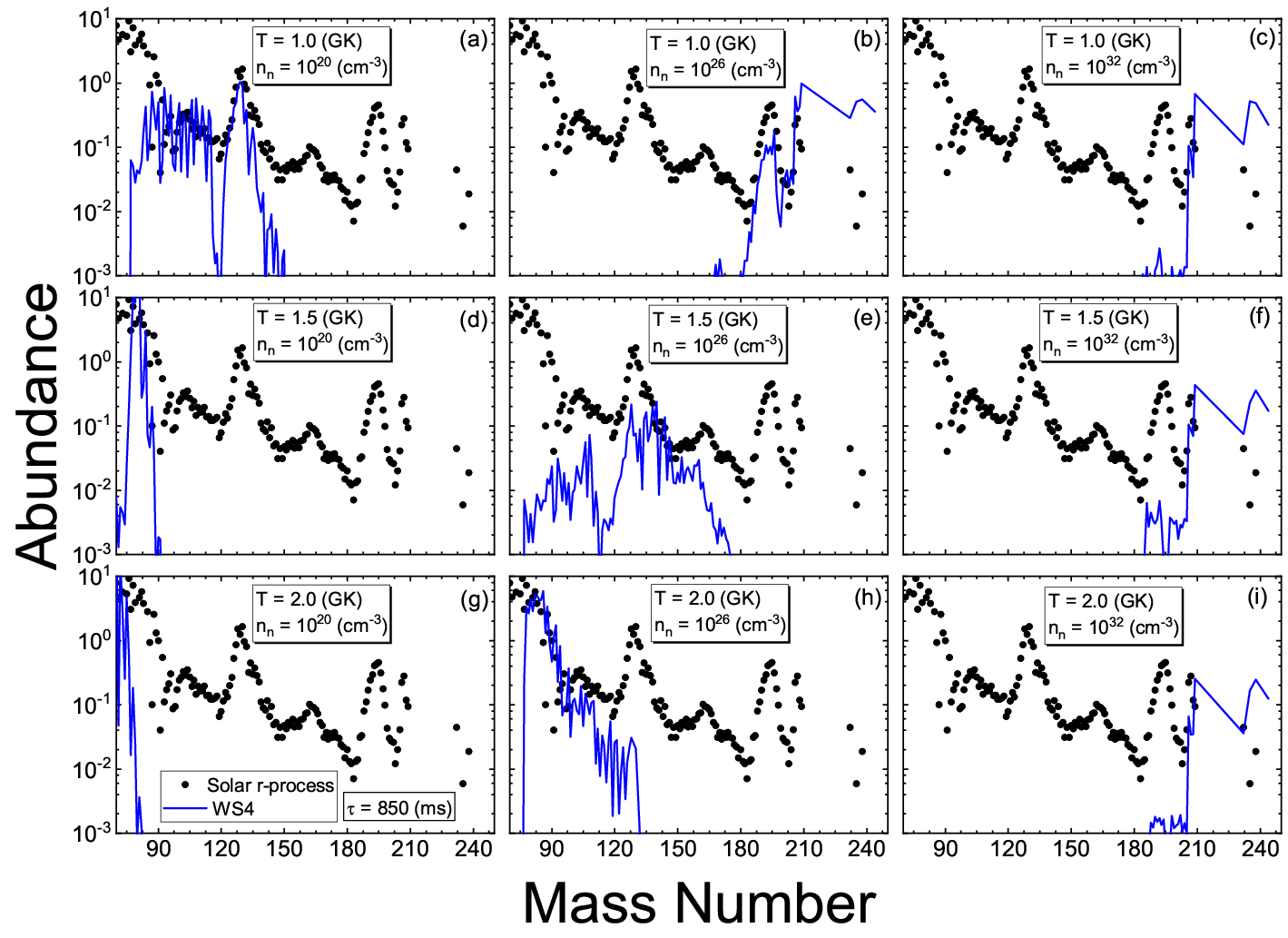}
  \caption{The $r$-process abundances under different astrophysical conditions, characterized by temperature ($T$) and neutron number density  ($n_n$), with a fixed irradiation time of $\tau=850~{\rm ms}$.
  }
\label{fig2}
\end{figure*}

The $r$-process abundances corresponding to different sets of astrophysical conditions (neutron number density $n_n$; temperature $T$) and a fixed irradiation time of $\tau=850~{\rm ms}$ are presented in Fig.~\ref{fig2}.
It can be observed that as both neutron densities increase and temperatures decrease, the $r$-process abundance distribution shifts toward the region of heavier nuclides.
This may indicate that the properties of near-neutron-drip-line nuclei are likely to be more important for determining the $r$-process abundances of heavier nuclides.

%----------------------------------------------------------------------------------------
%----------------------------------------------------------------------------------------

In order to quantitatively determine the relationship between the proximity of $r$-process paths to the neutron-drip line and astrophysical conditions, the average distance between an $r$-process path and the neutron-drip line is defined as follows:
\begin{equation} \label{L}
  L = \frac{1}{Z_f-Z_i}\sum_{Z=Z_i}^{Z_f}[ N_d(Z)- N_r(Z)],
\end{equation}
where $N_d(Z)$ and $N_r(Z)$ are the neutron numbers of the neutron-drip-line nucleus and the $r$-process path nucleus in the isotopic chain of a given proton number $Z$, respectively.
$Z_i$ and $Z_f$ are the picked initial and final proton numbers along the $r$-process path, which define the specific segment of the $r$-process path.
Note that each $r$-process path nucleus is characterized by a specific $(Z,N)$ pair, therefore, $Z_i$ ($Z_f$) also corresponds to a specific $N_i$ ($N_f$).
The smaller the distance $L$ is, the closer the $r$-process path is to the neutron-drip line.
When $L=0$, the $r$-process path lies directly on the neutron-drip line--a scenario that corresponds to extremely neutron-rich astrophysical conditions.

\begin{figure}[!ht]
  \centering
  \includegraphics[width=0.5\textwidth]{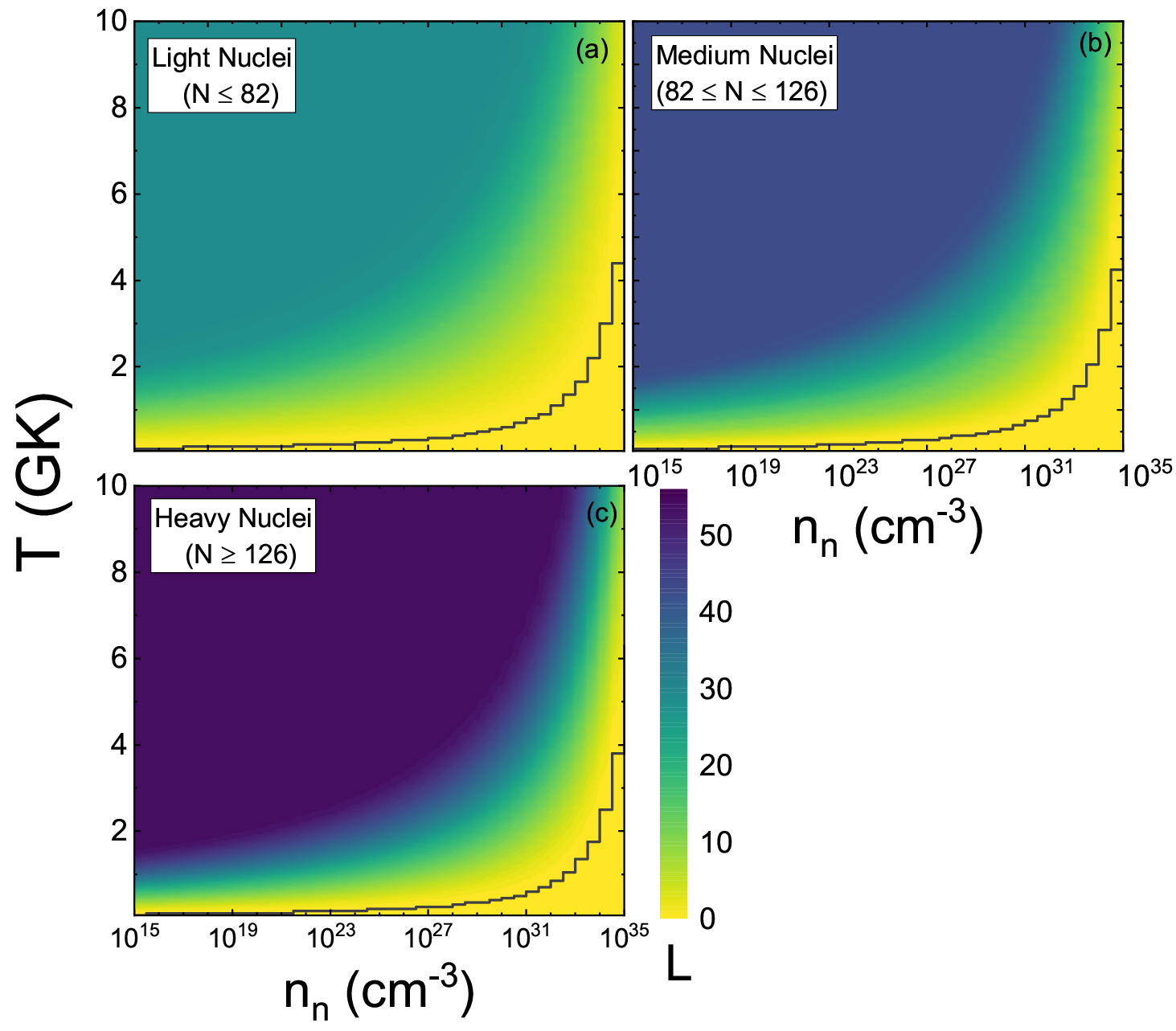}
  \caption{The averaged distances between $r$-process paths and the neutron-drip line under different astrophysical conditions, characterized by neutron number density $n_n$ and temperature $T$.
  Each $r$-process path is divided into three segments based on neutron number $N$: $N\leq82$ (light nuclei), $82\leq N\leq126$ (medium nuclei), and $N\geq126$ (heavy nuclei).
  The black lines denote the astrophysical conditions that first enable the $r$-process paths to reach the neutron-drip line.}
\label{fig3}
\end{figure}

The averaged distances between $r$-process paths and the neutron-drip line under different astrophysical conditions are illustrated in Fig.~\ref{fig3}.
Each $r$-process path is separated into three segments: with $N\leq82$ for light nuclei, $82\leq N\leq126$ for medium nuclei, and $N\geq126$ for heavy nuclei.
As can be seen in Fig.~\ref{fig3}, the distance $L$ decreases monotonically as neutron density increase and temperature decrease, and the $r$-process paths reach the neutron-drip line when the neutron density and temperature reach certain values.
Note that, in the region of light nuclei, the $r$-process paths can get relatively closer to the neutron-drip line under conditions of lower neutron density and higher temperature.
This is because the neutron-drip line in the lighter nuclear region does not extend as far as in the heavier nuclear region.

%----------------------------------------------------------------------------------------
%----------------------------------------------------------------------------------------

\section{Impact of near-neutron-drip-line nuclei on $R$-process abundances}

The next step is to examine the impacts of nuclear properties of near-neutron-drip-line nuclei on the $r$-process. 
For this purpose, a sensitivity study similar to that in Refs.~\cite{Mumpower2015Phys.Rev.C,Mumpower2016Prog.Part.Nucl.Phys., Jiang2021Astrophys.J.} is conducted.

\begin{figure}[!htbp]
  \centering
  \includegraphics[width=0.5\textwidth]{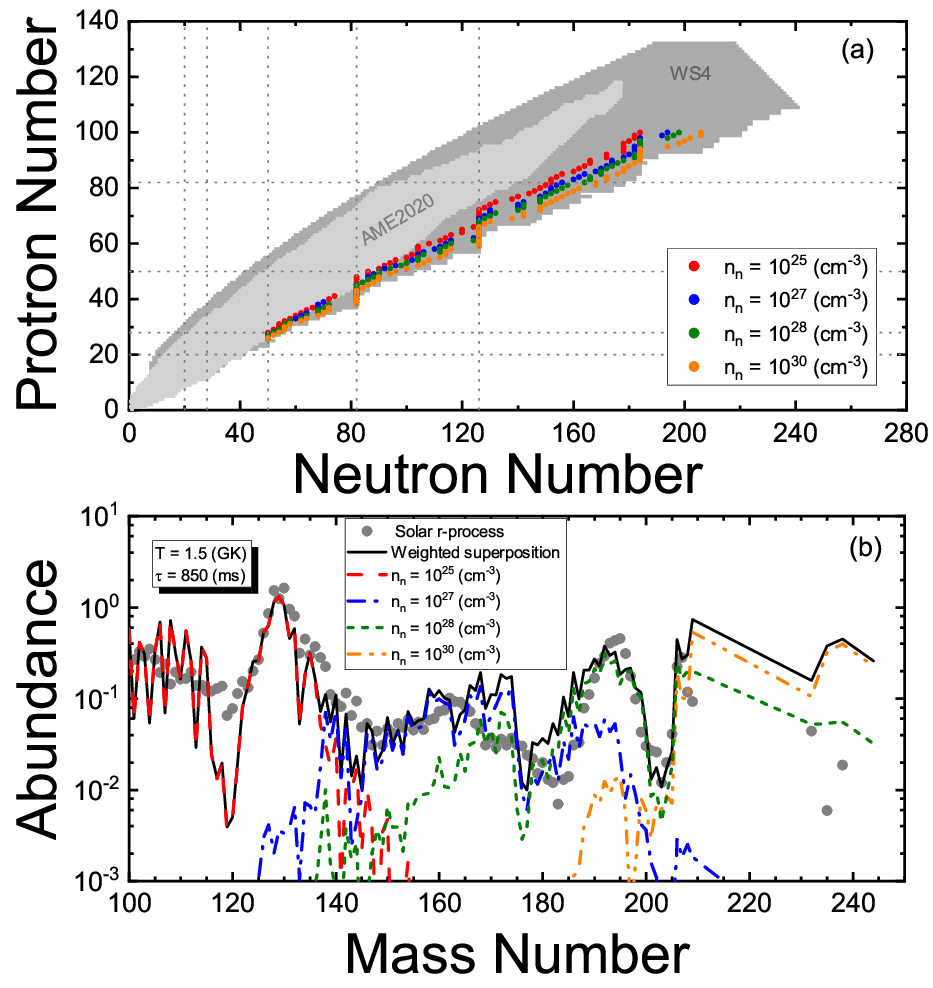}
  \caption{(a) $r$-process paths and (b) $r$-process abundances under four typical sets of astrophysical conditions. 
  The neutron number densities are $n_n = {10}^{25} \mathrm{~cm}^{-3}$, $n_n = {10}^{27} \mathrm{~cm}^{-3}$, $n_n = {10}^{28} \mathrm{~cm}^{-3}$, and $n_n = {10}^{30} \mathrm{~cm}^{-3}$, with fixed irradiation time $\tau  = 850~{\rm ms}$ and temperature $T = 1.5~{\rm GK}$.}
\label{fig4}
\end{figure}

Four sets of $r$-process simulations under astrophysical conditions with neutron number densities of $n_n = {10}^{25} \mathrm{~cm}^{-3}$, $n_n = {10}^{27} \mathrm{~cm}^{-3}$, $n_n = {10}^{28} \mathrm{~cm}^{-3}$, and $n_n = {10}^{30} \mathrm{~cm}^{-3}$, and a fixed irradiation time $\tau  = 850~{\rm ms}$ and temperature $T = 1.5~{\rm GK}$, are selected as typical cases that represent astrophysical conditions corresponding to the abundances of different mass regions.
$r$-process simulations with these conditions and unvaried nuclear physics inputs are regarded as baseline simulations.
Note that the astrophysical conditions adopted in this work are relatively extreme, which is intended to bring the $r$-process path close to the neutron-drip line. Nevertheless, these conditions remain reasonable and fall within the scope of certain parameterized environments for supernova explosions~\cite{Farouqi2010Astrophys.J.} and neutron star mergers~\cite{Lippuner2015Astrophys.J.} in the dynamic $r$-process studies.
The $r$-process paths and $r$-process abundances corresponding to these four sets of astrophysical conditions are presented in Fig.~\ref{fig4}.
It can be observed from Fig. \ref{fig4} (a) that the $r$-process paths of these four cases under the selected conditions are generally close to the neutron-drip line.
Note that not only the nuclei on the $r$-process paths can affect the $r$-process abundances, but also those close to the $r$-process path can affect the $r$-process abundances to varying extents.
In this regard, the properties of near-neutron-drip-line nuclei could, in principle, potentially affect the results of all these four sets of simulations.
The $r$-process simulations under these four cases correspond to the abundances of the $A\sim130$ peak, the rare-earth isotopes, the $A\sim 195$ peak, and superheavy radioactive nuclei respectively.
This will be helpful for analyzing how the properties of near-neutron-drip-line nuclei affect the $r$-process abundances of nuclei in different mass regions.
As can be seen in Fig.~\ref{fig4} (b), the weighted superposition of these four sets of $r$-process abundances can reproduce the Solar $r$-process abundances quite well.
This means that the current sets of simulations represent reasonably some realistic scenarios for $r$-process studies to reproduce the observed $r$-process abundance.

%----------------------------------------------------------------------------------------
%----------------------------------------------------------------------------------------

\begin{figure}[!htbp]
  \centering
  \includegraphics[width=0.5\textwidth]{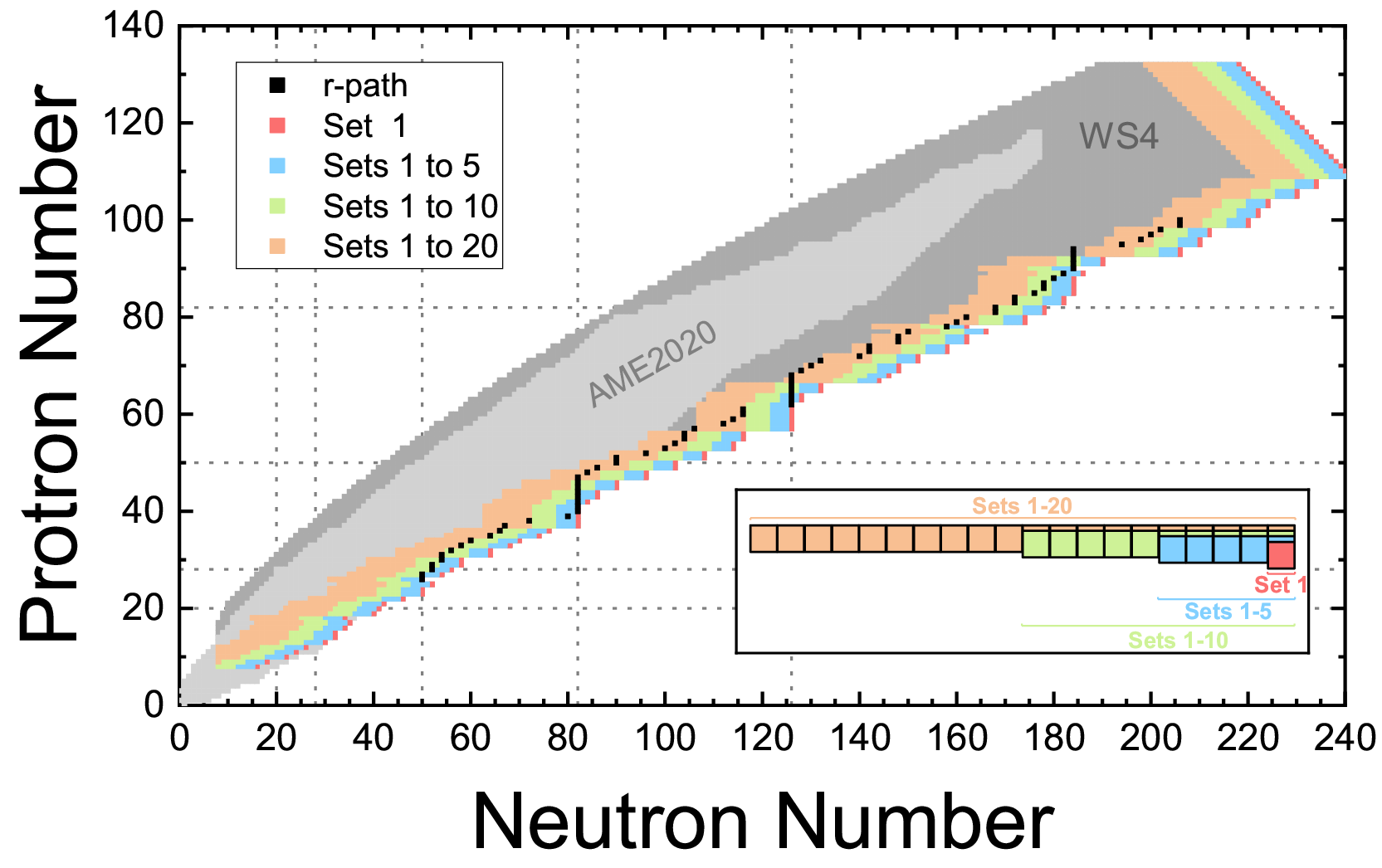}
  \caption{Illustrations of the different nuclear collections varied in the sensitivity studies: set 1, sets 1 to 5, sets 1 to 10, and sets 1 to 20. 
   Each collection consists of neutron-rich nuclei classified by their distance relative to the neutron-drip line of their respective isotopic chains.
   The $r$-process path is obtained by weighted superposition of the four paths in Fig.~\ref{fig4} (a).
   The inset subplot shows the detailed partitioning of the 20 most neutron-rich nuclei in each isotope chain into each collection.}
\label{fig5}
\end{figure}

\begin{figure}[!htbp]
  \centering
  \includegraphics[width=0.5\textwidth]{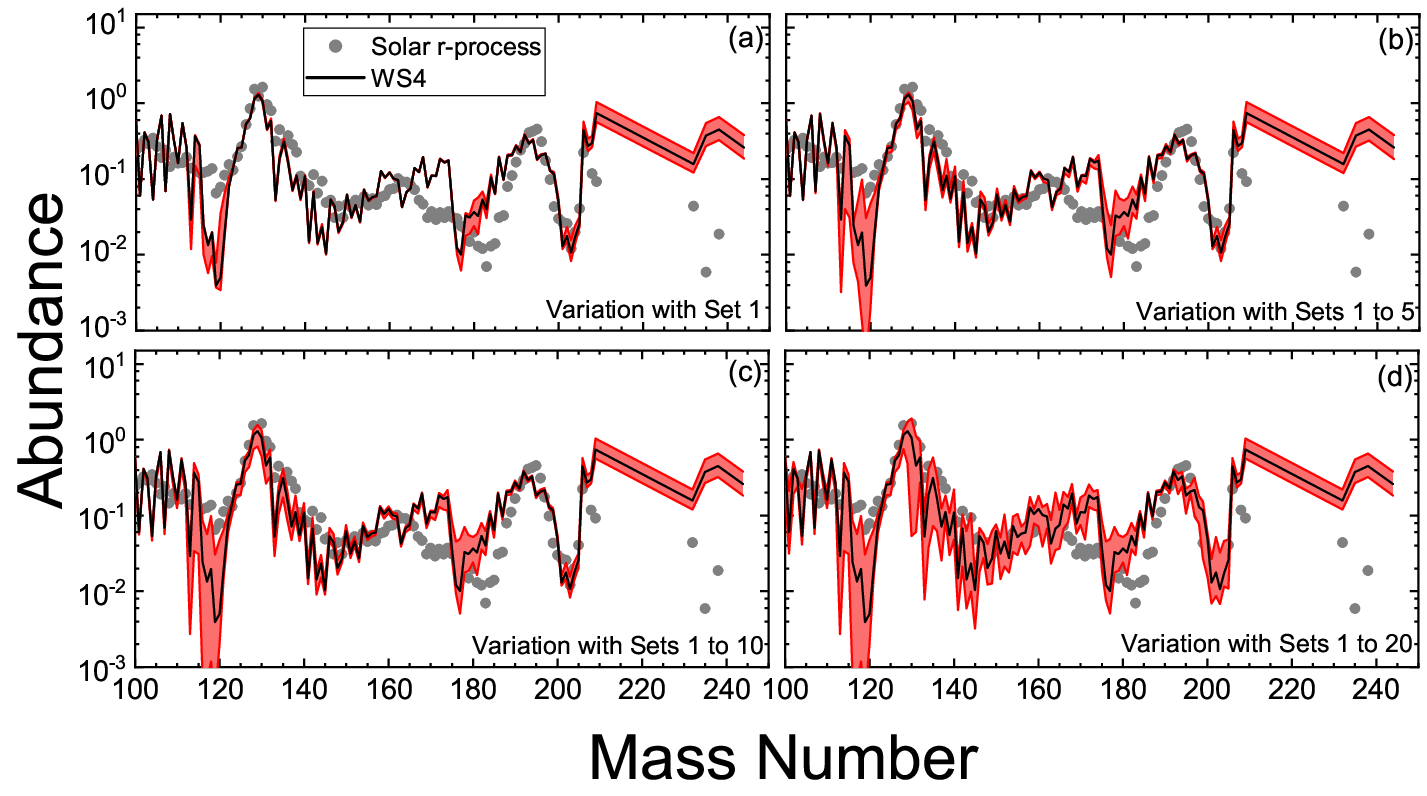}
  \caption{Variations of $r$-process abundances corresponding to the nuclear mass variations in the sensitivity studies. 
  Each subplot corresponds to the nuclear collection illustrated in the matching collection of Fig.~\ref{fig5}: (a) variations for set 1 nuclei, (b) variations for sets 1 to 5 nuclei, (c) variations for sets 1 to 10 nuclei, and (d) variations for sets 1 to 20 nuclei.
  The bands in each subplot represent the range of abundance variations caused by $\Delta M=\pm 0.5~{\rm MeV}$ nuclear mass variations.}
\label{fig6}
\end{figure}

To perform the sensitivity study on how nuclei near the neutron-drip line affect $r$-process abundances, the following steps are implemented:
First, the neutron-rich nuclei in each isotopic chain are classified into different sets according to their distances relative to the neutron-drip nucleus of that isotopic chain.
Figure~\ref{fig5} shows the collections of nuclei in set 1, sets 1 to 5, sets 1 to 10, and sets 1 to 20, respectively.
Sensitivity studies are then carried out by varying each predicted nuclear mass in these collection (for different numbers of sets of nuclei) respectively with $\Delta M=\pm 0.5~{\rm MeV}$ in the $r$-process simulations.
Note that only the upper and lower bounds of the nuclear mass variations are considered.
The $\beta$-decay half lives are varied consistently with the mass variations by imposing different $Q_{\beta}$ values in the empirical formula~\cite{Zhou2017Sci.ChinaPhys.Mech.Astron.}.
Subsequently, a number of resultant $r$-process abundance patterns are obtained, and they form a band as shown in each subplot of Fig.~\ref{fig6}.

As can be seen from Fig.~\ref{fig6}, the variations in the nuclear masses of near-neutron-drip-line nuclei can directly lead to significant variations in the abundances of superheavy nuclei, i.e., lead and actinide nuclei.
The abundances of these nuclei are mainly produced under the astrophysical condition with $n_n = {10}^{30} \mathrm{~cm}^{-3}$ (see Fig.~\ref{fig4} (b)), and the $r$-process path associated with this condition is mostly close to the neutron-drip line, as can be seen in Fig.~\ref{fig4} (a).

One can also see from Fig. \ref{fig6} that the abundance variations are obvious in the regions with $A=110-125$, $A=175-185$, and $A=200-205$,  just before respective abundance peaks. 
The reasons for the variations with $A=110-125$ and $A=175-185$ are the same, which are twofold.
One reason arises from differences in abundance magnitudes between these nuclei and the peak nuclei.
The abundances of these nuclei are lower by more than one order of magnitude compared with the peak abundances; therefore, even small variations in the abundances of peak nuclei can lead to significant variations in the abundances of these nuclei.
This reasoning also applies to the $A=200-205$ region.
The other reason involves neutron shell effects across different isotopic chains.
For example, the $r$-process path nuclei associated with the $N=82$ neutron shell holds for isotope chains with $Z=40-47$.
The abundances accumulated in these nuclei contribute to the abundances of nuclei in and around the $A=130$ $r$-process peak.
Since the nuclei with smaller $Z$, e.g., $Z=40$, are closer to the neutron-drip line, they are affected more significantly than the others in the sensitive studies.
This leads to larger abundance variations in nuclei with $A=122$ and below, due to $\beta$-delay neutron emissions, and this is the reason for the abundance variations in the $A=110-125$ region.
Similar mechanisms also cause the obvious abundance variations in the $A=175-185$ region, which relate to the $N=126$ neutron shell.
Another reason for the nuclear mass variations in the $A=200-205$ region is the same as that for lead and actinide nuclei: the corresponding $r$-process path is closest to the neutron-drip line.

One important indication is that the abundances of the lanthanide nuclides, especially those in the $r$-process rare-earth peak, are affected only when the number of variation sets reaches 20. 
This is good news for studies that aim to understand the origin of the rare-earth peak, as it may not be strongly affected by the large uncertainties of nuclei close to the neutron-drip line.
It should also be noted that the $r$-process peaks at $A=130$ and $A=195$ are not affected by variations in the nuclear masses of nuclei close to the neutron-drip line.
This is reasonable, as it is well known that $r$-process simulations based on different nuclear mass models, even those that differ in the locations of the neutron-drip line, always yield these two $r$-process peaks consistently.

%----------------------------------------------------------------------------------------
%----------------------------------------------------------------------------------------

%----------------------------------------------------------------------------------------
%----------------------------------------------------------------------------------------

It is also important to identify which nuclei near the neutron-drip line are important for the $r$-process abundances.
To quantify the abundance variations caused by varying the mass of each nucleus, the abundance deviation $\Delta Y$ is introduced
\begin{equation}
  \Delta Y = \frac{1}{N} \sum_{A=100}^{244}|\log Y_{\rm max}(A) - \log Y_{\rm min}(A)|,
\end{equation}
where, $Y_{\rm max}(A)$ ($Y_{\rm min}(A)$) is the largest (smallest) abundance obtained from $r$-process simulations with baseline inputs, baseline inputs with $\Delta M=+0.5~{\rm MeV}$ for a specific nucleus, and baseline inputs with $\Delta M=-0.5~{\rm MeV}$ for that nucleus.

%----------------------------------------------------------------------------------------
\begin{figure}[!htbp]
  \centering
      \includegraphics[width=0.5\textwidth]{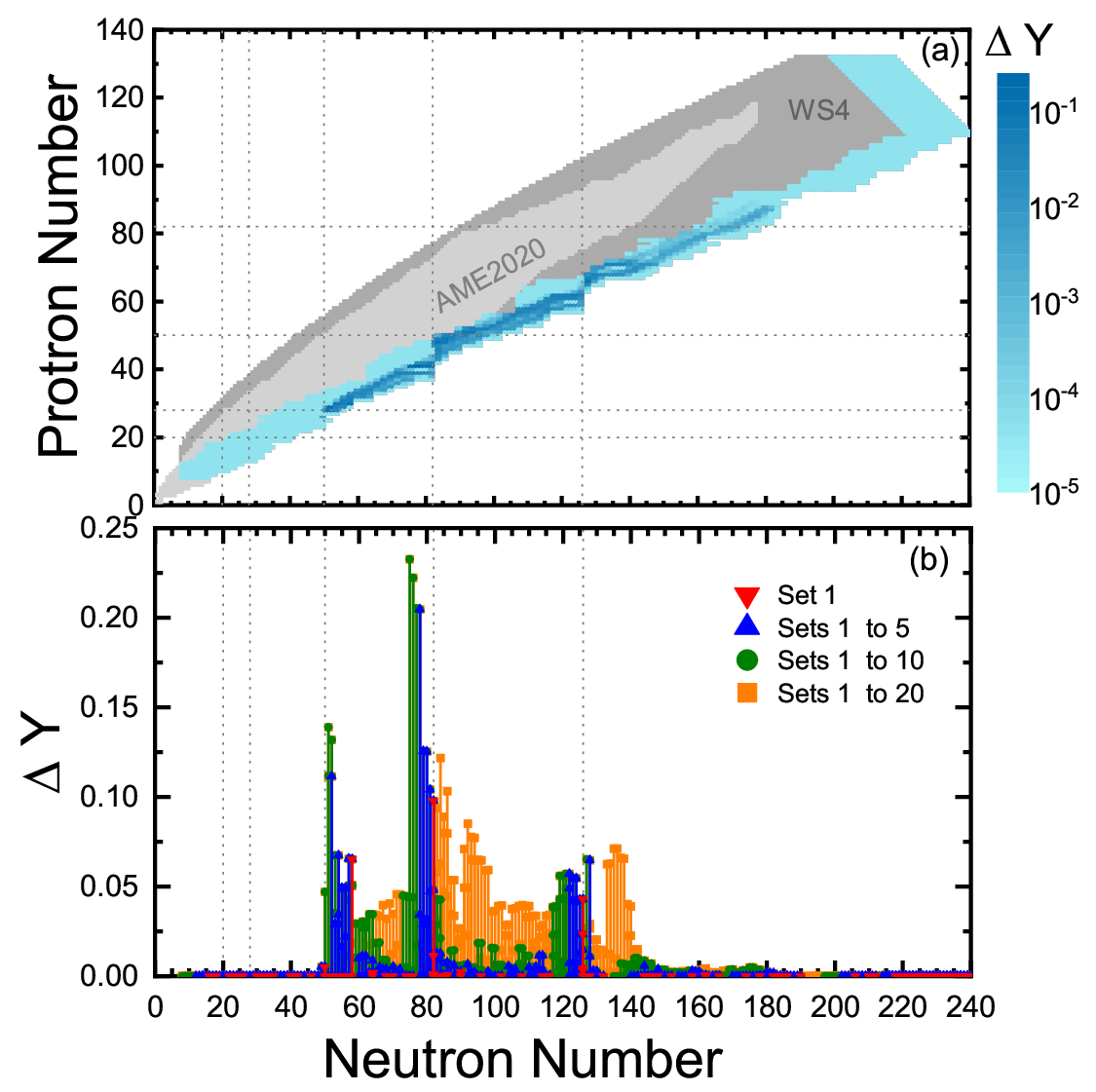}
  \caption{(a) Nuclear chart presenting the $r$-process abundance deviations $\Delta Y$ caused by varying nuclear mass of each single nucleus.
  (b) The abundance deviations $\Delta Y$ are plotted as a function of neutron number to have a clear visualization with colors label nuclei in different sets.
  Note that panel (b) is just another representation of $\Delta Y$ in panel (a), it does not sum over $\Delta Y$ caused by nuclei with the same $N$.}
\label{fig7}
\end{figure}
%----------------------------------------------------------------------------------------

The $r$-process abundance deviations caused by varying the mass of each single nucleus are presented in Fig.~\ref{fig7}.
As can be seen from Fig.~\ref{fig7} (a), the nuclei that obviously impact $r$-process abundances are mainly distributed in the region of $25\leq Z\leq 90$ and $50\leq N\leq 180$.
In the classical $r$-process model, one uses iron as the seed nucleus, so lighter nuclei are ignored.
It is encouraging to find that nuclei heavier than $A=270$ cause relatively small uncertainties in $r$-process abundances; otherwise, one could lead to large uncertainties in understanding the $r$-process, as there are still significant uncertainties in theoretically describing these heavy neutron-rich nuclei.
However, this could also be due to the fact that fission is not included in the classical $r$-process model employed here. 
The masses of $A>270$ nuclei can affect their fission rates, which in turn can influence $r$-process abundances.
As is clearly shown in Fig.~\ref{fig7} (b), when varying the masses of nuclei from set 1 (neutron-drip line nuclei) to set 20 (neutron-rich nuclei), the nuclei that significantly impact r-process abundances first appear around the neutron magic numbers $N = 50$ with $Z \approx 28$, $N = 82$ with $Z \approx 41$, and $N = 126$ with $Z \approx 66$. It is then extend to other regions, mainly around the rara-earth isotopes.
This indicates that nuclei around neutron magic numbers are important for the $r$-process, even in the near-neutron-line region.
Note that a large deviation can be observed between $N=82$ and $N=126$ in Fig.~\ref{fig7} (b), which is related to the lanthanide region. This region is of great interest to both nuclear physics and astrophysics: it is the nuclear shape-transition region, and the origin of the rare-earth $r$-process abundance peak in this region remains unresolved~\cite{Surman1997Phys.Rev.Lett., Goriely2013Phys.Rev.Lett.}. 
Since the deviation here is quite significant for Sets 11 to 20, a more accurate description of nuclei located over a dozen nucleons from the drip line is required to reveal the origin of the rare-earth $r$-process abundance.

\section{Summary}

In summary, the role of near-neutron-drip-line nuclei in the $r$-process is studied with the classical $r$-process model.
By performing $r$-process simulations under different astrophysical conditions $(T,n_n)$, it is found that lower temperatures and higher neutron densities drive $r$-process paths to approach the neutron-drip line and shift abundance distributions toward heavier nuclides.
Quantitatively, the distance between the $r$-process path and the neutron-drip line decreases monotonically with increasing neutron density and decreasing temperature across all mass regions.

A weighted superposition of four sets of $r$-process simulations under different neutron number densities reproduces Solar $r$-process abundances quite well.
Based on this baseline, a sensitivity study further demonstrates that variations in the nuclear masses of near-neutron-drip-line nuclei significantly impact the abundances of superheavy nuclei and lead to obvious abundance variations in the $A=110-125$, $A=175-185$, and $A=200-205$ regions.
In contrast, the $r$-process rare-earth peak and the $A=130,195$ peaks remain largely unaffected.
The nuclei that obviously impact $r$-process abundances are mainly distributed in the region of $25\leq Z\leq 90$ and $50\leq N\leq 180$, with the nuclei around neutron magic numbers found to be particularly important for the $r$-process, even in the near-neutron-drip-line region.

These results highlight the need for precise constraints on the properties of near-neutron-drip-line nuclei to advance our understanding of $r$-process nucleosynthesis.

\begin{acknowledgments}
This work was supported by the National Natural Science Foundation of China (Grants No. 12405134, No. 12435006, No. 12141501, No. 12475117), the National Key R\&D Program of China (Contract No. 2024YFE0109803, No. 2024YFA1612600), the State Key Laboratory of Nuclear Physics and Technology, Peking University (Grant No. NPT2023ZX03, No. NPT2025KFY02), and the National Key Laboratory of Neutron Science and Technology (Grant No. NST202401016).
\end{acknowledgments}

\bibliography{paper}

\end{document}